%% file: main.tex
\documentclass[conference,letterpaper]{IEEEtran}
\IEEEoverridecommandlockouts

\usepackage{cite}
\usepackage{amsmath,amssymb,amsfonts}
\usepackage{algorithmic}
\usepackage{capt-of}
\usepackage{graphicx}
\usepackage{textcomp}
\usepackage{xcolor}
\usepackage{url}
\graphicspath{{imgs/}}

\def\BibTeX{{\rm B\kern-.05em{\sc i\kern-.025em b}\kern-.08em
    T\kern-.1667em\lower.7ex\hbox{E}\kern-.125emX}}

\title{SNAS: A Multi-Layer Defense-in-Depth Architecture for Secure Egress in Sandboxed Workloads}

\author{
\IEEEauthorblockN{
Niranjan Kumar Sharma, S Muralidhar, Samy Boshra-Riad, Mike Halcrow,\\
Yuxiong He, Nitya Kumar Sharma, Shawn Xia, Haowei Yu,\\
Elliott Brossard, Derek Denny-Brown, Choden Konigsmark,\\
Bhanu Prakash, Brandon Baker, Andong Zhan
}
\IEEEauthorblockA{
\textit{Snowflake Inc.}\\
USA
}
}

\begin{document}
\pagestyle{empty}

\IEEEpubid{\makebox[\columnwidth]{\footnotesize\copyright~2026 IEEE.
Accepted at the 53rd IEEE/IFIP DSN 2026.
Personal use is permitted; other uses require IEEE permission.\hfill}%
\hspace{\columnsep}\makebox[\columnwidth]{}}

\maketitle

\begin{abstract}
Snowpark enables data engineering and AI/ML workloads in Snowflake by executing user-defined functions in secure sandboxes. Many of these workloads require external connectivity to access cloud APIs, external databases, or feature stores, creating a dependability challenge: how to provide transparent network access while preserving strict multi-tenant isolation and resource fairness. This paper presents Secure Network Access in Snowpark (SNAS), a production architecture for secure external communication from sandboxed workloads. SNAS combines Extended Berkeley Packet Filter (eBPF) packet filtering, Generic Network Virtualization Encapsulation (GENEVE) overlay networks, and distributed egress proxies for policy-driven egress control with low overhead. We describe the design, deployment, and measured production behavior of SNAS, including an eBPF-based bandwidth limiter using the Earliest Departure Time (EDT) algorithm, dual-tier policy enforcement, and safeguards for connection limiting and port exhaustion. SNAS is deployed across all Snowflake regions and supports large-scale production workloads including petabyte-scale data transfer and latency-sensitive external integrations.
\end{abstract}


\input{chapters/introduction}
\input{chapters/related_work}
\input{chapters/background}
\input{chapters/snas_architecture}
\input{chapters/challenges}
\input{chapters/evaluation}
\input{chapters/deployment}
\input{chapters/conclusion}

\section*{Acknowledgment}
SNAS is the result of dedicated engineering effort across multiple teams at Snowflake. We thank the Snowpark, Infrastructure, and Cloud Services teams for their contributions. We are grateful to our customers whose feedback shaped the system's evolution.

\bibliographystyle{IEEEtran}
\bibliography{bib/references}

\end{document}

%% file: chapters/introduction.tex
\section{Introduction}

Snowflake~\cite{cite1} pioneered the separation of compute and storage in data warehousing through an elastic, multi-tenant architecture. Snowpark~\cite{cite2} extended Snowflake beyond SQL to support data engineering and AI/ML workloads in Python, Java, and other languages through secure sandboxing. In operating Snowpark at scale, we observed a recurring practical challenge: customers increasingly need direct, policy-controlled access to external services while still expecting the isolation and dependability guarantees of a managed multi-tenant platform. Modern data pipelines depend on external integration. Customers building cloud data connectors often require native API access using gRPC or custom protocols. Machine learning workflows need real-time feature enrichment from external caching layers. Geospatial analytics must call geocoding APIs. Database federation requires JDBC connectivity to external sources. Supporting these use cases in production required moving beyond the constraints of the prior External Functions~\cite{cite3} mechanism (which was limited to HTTPS+JSON and involved serialization overhead) toward a transparent, multi-protocol access model. This paper focuses on the industrial design, deployment, and operational behavior of that solution. To address these limitations while maintaining Snowflake's security standards, we established three design objectives.

\IEEEpubidadjcol

\begin{enumerate}
  \item \textbf{Transparency:} The architecture enables standard POSIX socket operations, allowing unmodified usage of third-party libraries (e.g., Python \texttt{requests}) without proprietary SDKs.

  \item \textbf{Security:} Strict egress control with a zero-trust, defense-in-depth architecture, featuring multiple validation layers to prevent data exfiltration even under zero-trust assumptions.

  \item \textbf{Performance:} To ensure fair allocation in multi-tenant environments, we maintain sub-millisecond overhead for petabyte-scale workloads by using eBPF-based resource management. Our solution leverages eBPF for programmable packet filtering, GENEVE tunnels for traffic isolation, and distributed egress proxies for centralized enforcement. We implement bandwidth limiting using the Earliest Departure Time (EDT) algorithm~\cite{cite5}, a hybrid approach to DDoS prevention using sliding-window-based connection limiting~\cite{cite4}, and management to prevent port exhaustion. The system has been deployed across all Snowflake regions.
\end{enumerate}

\textbf{Threat Model.} SNAS is designed for a multi-tenant production environment with three main concerns: malicious user code attempting unauthorized external access, potentially compromised compute nodes, and noisy-neighbor workloads that consume disproportionate bandwidth. Accordingly, SNAS treats sandbox-originated traffic as untrusted and compute nodes as potentially compromised.

The primary contributions of this work are:

\begin{enumerate}
    \item Multi-Layer, Defense-in-Depth Egress Architecture: We present the production architecture used for secure multi-tenant egress in Snowpark, combining eBPF packet filtering at the sandbox level, GENEVE overlays for traffic isolation, and a distributed egress proxy tier for independent, centralized policy enforcement.
    \item eBPF-Based Resource Management in Practice: We describe the use of the Earliest Departure Time (EDT) algorithm~\cite{cite5} via eBPF for precise, low-overhead bandwidth limiting, along with a hybrid sliding window connection limiter to mitigate abuse and prevent resource exhaustion.
    \item Transparent Support for Developer Workloads: Our system enables standard networking libraries and common protocols to function from within the sandbox without modification, proxy configuration, or Snowflake-specific APIs.
    \item Validation at Global Production Scale: We report measured deployment and operational experience from SNAS across all Snowflake regions, including daily workload volume, large-scale data transfers, and lessons learned from running the system in production.
\end{enumerate}

%% file: chapters/related_work.tex
\section{Related Work}
\label{sec:related-work}

Providing secure, multi-tenant network access from sandboxed environments is a significant challenge. Our work builds upon and contrasts with several key areas of research and existing technologies. 

\begin{enumerate}
    \item Cloud-Native Networking with eBPF. Projects like Cilium~\cite{cite8} validate eBPF's utility as a general-purpose Container Network Interface (CNI) for Kubernetes, providing broad pod-to-pod and ingress/egress security. In contrast, SNAS is a specialized, egress-only architecture, tightly integrated with the data platform's control plane. General CNIs enforce cluster-scoped policies defined by operators, trust the node's policy agent, and require distributed policy stores. SNAS policies are instead compiled per-job by the platform's control plane, cryptographically signed, ephemeral, and validated independently at two tiers. This provides defense against compute node compromise, a threat model beyond standard CNI assumptions. In addition, SNAS integrates EDT-based bandwidth limiting directly into the enforcement path to provide per-tenant fairness that general-purpose CNI solutions do not natively offer.
    \item Serverless Sandboxing and Isolation. Technologies like gVisor~\cite{cite13} provide essential compute and memory isolation for serverless workloads. These solutions, however, do not inherently solve for secure, managed external networking. Our work is complementary, providing the missing policy-driven, resource-managed egress architecture for these types of isolated compute environments.
    \item Snowflake's prior solution, External Functions~\cite{cite3}, is a proxy-based system designed for HTTPS+JSON request-response calls. While effective for its original scope, its architecture introduced protocol constraints (preventing use cases requiring gRPC or JDBC), serialization overhead, and required additional, user-managed cloud configuration. SNAS directly addresses these areas by enabling transparent, multi-protocol network access with minimal overhead.
    \item Network Access Control and Firewalling. Network-level policy enforcement has an extensive research history. Wool~\cite{cite20} demonstrated that firewall misconfigurations are prevalent even in well-managed enterprises, motivating automated policy compilation. FIREMAN~\cite{cite21} introduced rule-conflict analysis. SDN approaches~\cite{cite22} decouple policy specification from enforcement. Unlike traditional firewalls operating on static rulesets, SNAS policies are job-scoped and ephemeral—compiled per-job and automatically expired after completion.
    \item Traffic Shaping and Multi-Tenant Isolation. Fair queuing algorithms~\cite{cite28} provide foundational bandwidth allocation mechanisms. The EDT algorithm~\cite{cite5} assigns per-packet departure timestamps for pacing. FQ-CoDel~\cite{cite29} addresses bufferbloat. SNAS applies EDT in a production multi-tenant context: per-tenant bandwidth limiting implemented entirely in eBPF with policy-keyed rate state. Network virtualization overlays such as VXLAN~\cite{cite26} and GENEVE~\cite{cite9} provide L2-over-L3 encapsulation; SNAS uses GENEVE not merely for isolation but as a policy transport mechanism via TLV metadata. Firecracker~\cite{cite12} and SAND~\cite{cite27} address serverless compute isolation and communication respectively, but do not address managed external egress networking.
\end{enumerate}

%% file: chapters/background.tex
\section{Background}
\label{sec:background}

\subsection{Snowflake Architecture}

\begin{figure}[ht!]
  \centering
  \includegraphics[width=0.9\linewidth]{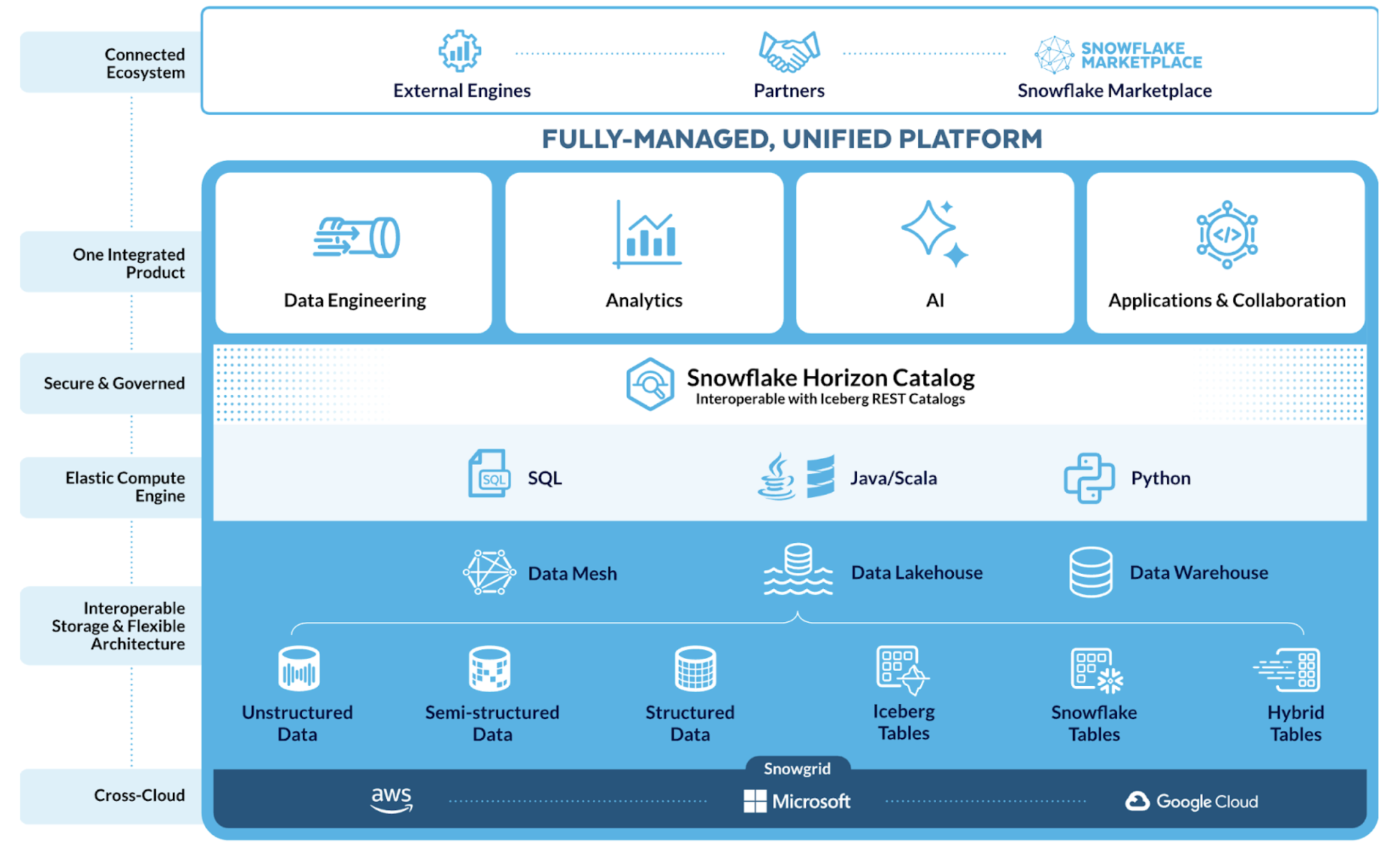}
  \caption{Snowflake architecture.}
  \label{fig:snowflake-arch}
\end{figure}

Snowflake's architecture~\cite{cite1}, as a software-as-a-service (SaaS) solution, employs a multi-cluster shared data model that fundamentally decouples the data storage and compute layers to facilitate independent, elastic resource scaling, as illustrated in Fig.~\ref{fig:snowflake-arch}. This architectural choice yields significant operational advantages. Specifically, the system operates as a pure SaaS platform, abstracting away all infrastructure provisioning and management responsibilities from the user. This fundamental reduction in operational overhead minimizes friction between data engineering and IT/infrastructure teams, leading to accelerated solution deployment.

\subsection{Snowpark Execution Environment}

Snowpark~\cite{cite2} enables Python, Java, and Scala execution within Snowflake. Stored procedures execute procedural logic; UDFs transform row or batch data, with vectorized UDFs using Pandas DataFrames~\cite{cite6}; table and aggregate functions support table generation and custom aggregation.

\subsection{Snowpark Security Model}

Snowpark uses gVisor~\cite{cite14} for multi-tenant isolation. Its Sentry intercepts and emulates most Linux system calls in user space, reducing host-kernel exposure. Combined with Linux namespaces and cgroups, this provides the isolation needed for controlled external access.

\subsection{eBPF Technology}

eBPF~\cite{cite7} enables safe Linux kernel extension without kernel modules. The verifier enforces safety and termination, JIT compilation provides near-native performance, and programs attach to hooks such as XDP, TC, and socket filters to modify, redirect, or drop packets. Cilium~\cite{cite8} demonstrates eBPF's production use in cloud-native networking.

%% file: chapters/snas_architecture.tex
\section{SNAS Architecture}
\label{sec:snas-architecture}

\subsection{User Model and Network Policy Definition}

\begin{figure}[ht!]
  \centering
  \includegraphics[width=0.9\linewidth]{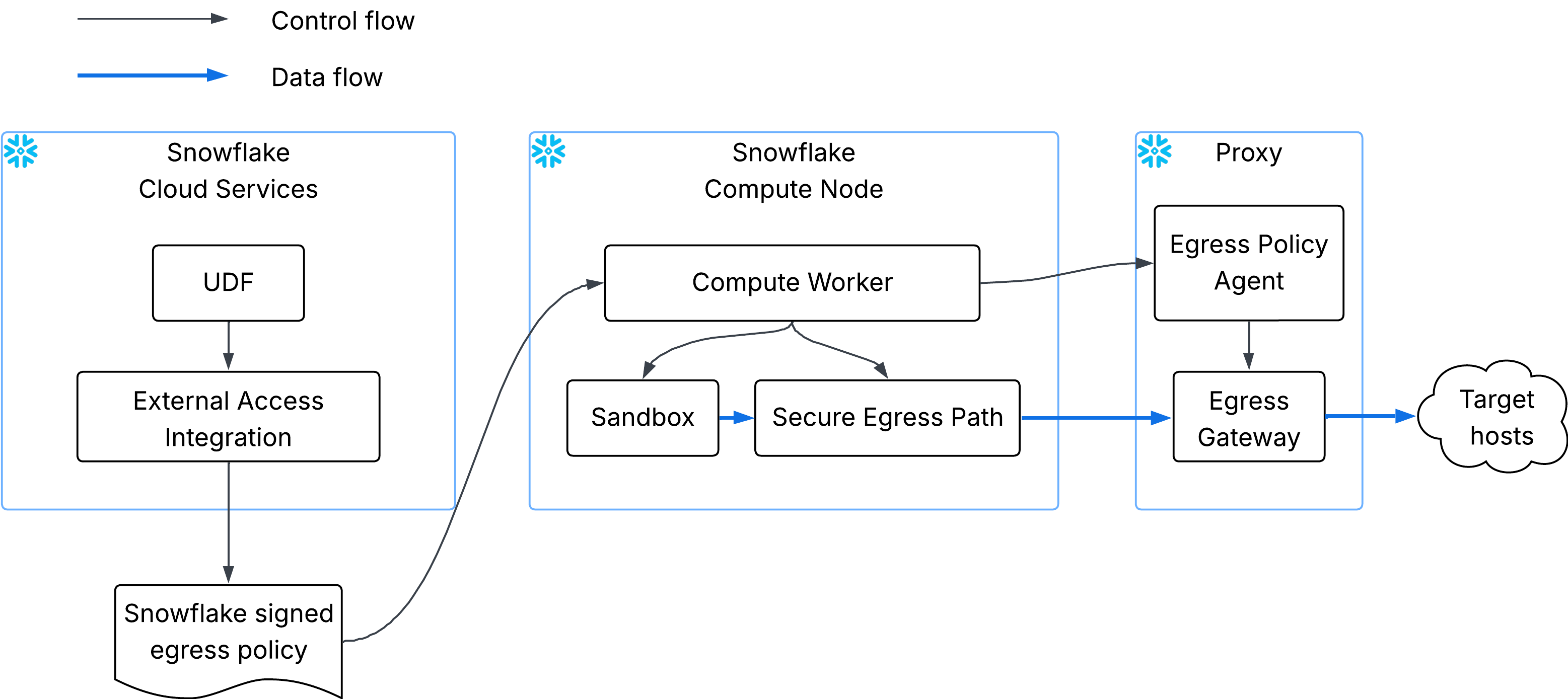}
  \caption{SNAS architecture overview.}
  \label{fig:snas-arch}
\end{figure}

SNAS is controlled through Network Rules and External Access Integrations~\cite{cite16}. Rules define allowed FQDNs, hostnames, or IP ranges; integrations group rules and secrets into policies referenced by UDFs. For each job, Cloud Services authorizes access, resolves hostnames, compiles allowed IP:port tuples, signs the policy, and distributes it to warehouse nodes and egress proxies (Fig.~\ref{fig:snas-arch}).

\subsection{Overlay Network Architecture}

The core design is a software-defined overlay network creating an isolated egress path from sandboxes through egress proxies. The architecture comprises three tiers providing defense-in-depth security, detailed in Fig.~\ref{fig:snas-overlay}:

\begin{figure}[ht!]
  \centering
  \includegraphics[width=0.9\linewidth]{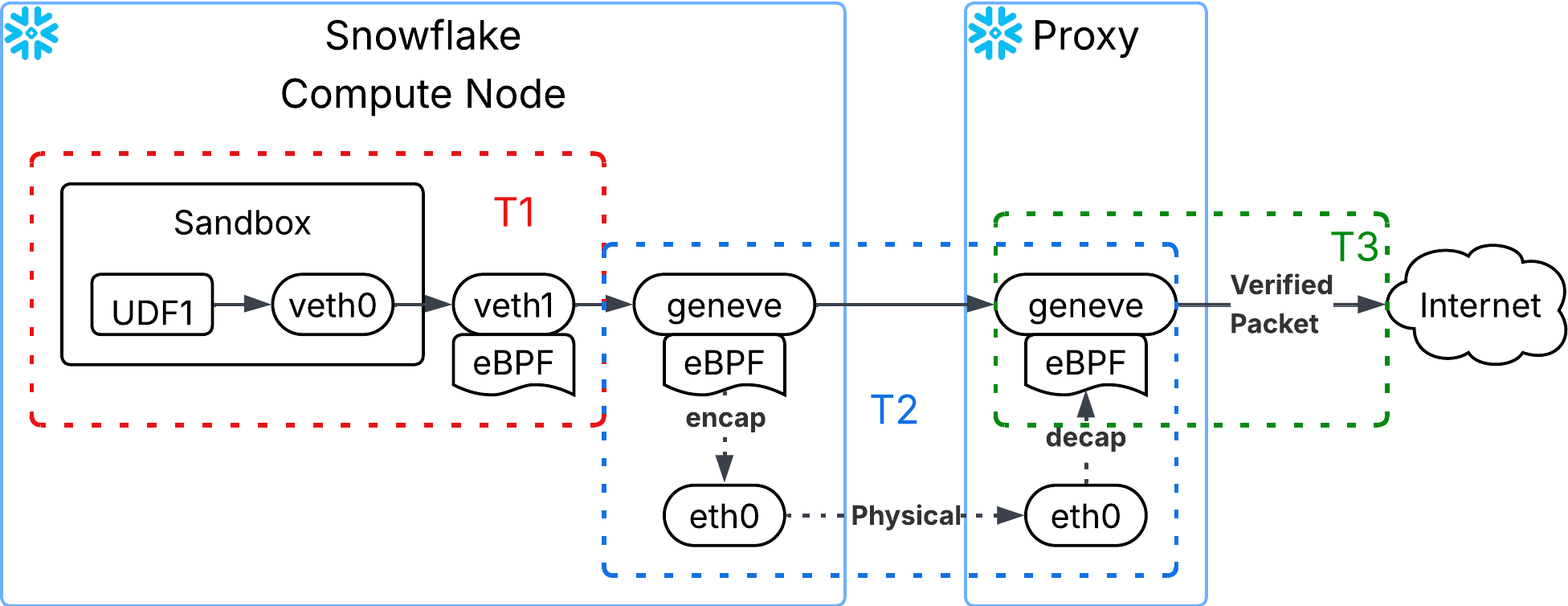}
  \caption{SNAS Overlay network.}
  \label{fig:snas-overlay}
\end{figure}

\begin{enumerate}

    \item \textbf{Sandbox to Compute Node.} Each sandbox gets a dedicated network namespace with a veth pair. One end of the pair appears as a standard Ethernet interface inside the sandbox, while the other end remains in the compute node namespace serving as the enforcement point. An eBPF program attached to the compute node side veth Traffic Control (TC) egress hook intercepts every packet, validating destination IP:port against the egress policy loaded in eBPF maps. Only authorized packets proceed; violations are logged and blocked immediately. This ensures the sandbox cannot access any IP:port not explicitly listed in its policy, including other sandboxes, Snowflake infrastructure, or arbitrary internet hosts. 
    \item \textbf{Compute Node to Egress Proxy.} Approved packets are encapsulated using GENEVE~\cite{cite9} (Fig.~\ref{fig:geneve}), embedding the policy ID in GENEVE metadata. The encapsulation changes the outer IP header destination to the egress proxy while preserving the original packet. GENEVE was selected for its UDP-based transport protocol enabling arbitrary destination IP setting, flexible Type-Length-Value (TLV) metadata format for embedding policy context, and status as an IETF standard with efficient kernel implementations. This provides transparent routing and stateless proxy operation—the policy ID provides all necessary context.

    \begin{minipage}{\linewidth}
      \centering
      \vspace{0.2cm}
      \includegraphics[width=0.9\linewidth]{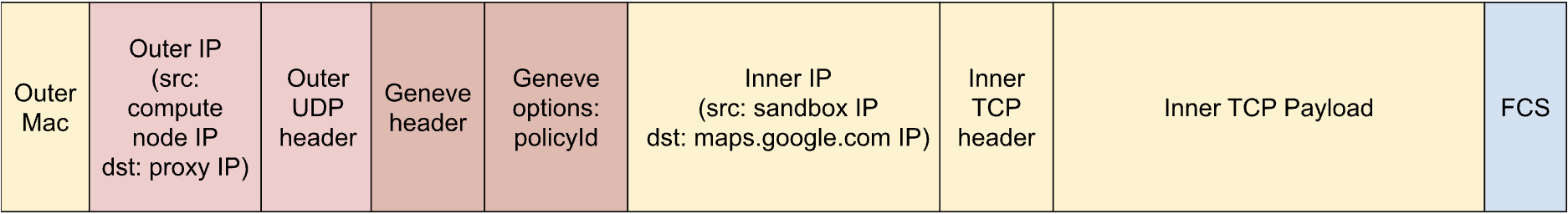}
      \captionof{figure}{GENEVE encapsulation.}
      \label{fig:geneve}
      \vspace{0.2cm}
    \end{minipage}
    
    \item \textbf{Egress Proxy to Internet.} Egress proxies decapsulate GENEVE packets, extract policy IDs, and perform secondary validation against cryptographically-signed egress policies that are compiled, signed, and distributed by Cloud Services~\cite{cite15}. This defense-in-depth prevents data exfiltration even if compute nodes are compromised. The defense-in-depth architecture is crucial, as the egress proxy validation provides a second, independent enforcement point, implementing a zero-trust model where it assumes no trust relationship with the compute nodes. Source Network Address Translation (SNAT) converts sandbox private IPs to proxy public IPs, with connection tracking for response routing. There is comprehensive logging for network activity oversight and security forensics.
\end{enumerate}

\subsection{Network Resource Limitations}
Enabling external access introduces resource consumption challenges addressed through comprehensive management mechanisms. 

\begin{enumerate}
    \item \textbf{Bandwidth Limiting.} We implement bandwidth limiting using the Earliest Departure Time (EDT) algorithm~\cite{cite5}. EDT assigns each packet a timestamp indicating earliest transmission time, providing superior performance versus traditional queuing disciplines through reduced latency, minimal packet loss, and O(1) complexity, as shown in Fig.~\ref{fig:edt-alg}. Our eBPF program attaches to the GENEVE device's TC egress hook. For each packet, the program retrieves bandwidth configuration from an eBPF map indexed by policy ID, calculates transmission schedule based on packet size and configured rate, sets the EDT timestamp on the sk\_buff (socket buffer) structure instructing the kernel when the packet may depart, and updates the last departure time in the eBPF map for the next packet in the flow. This approach leverages the kernel's Fair Queue (FQ) qdisc which respects EDT timestamps, pacing packets without queuing delays. The timing wheel data structure provides O(1) insertion and deletion complexity, maintaining efficiency even under high packet rates. Limits on bandwidth are configurable per compute node and documented in the egress policies. Default limits are set, scaling with warehouse types to ensure a balance between workload performance and infrastructure protection.

    \begin{minipage}{\linewidth}
      \centering
      \vspace{0.2cm}
      \includegraphics[width=0.9\linewidth]{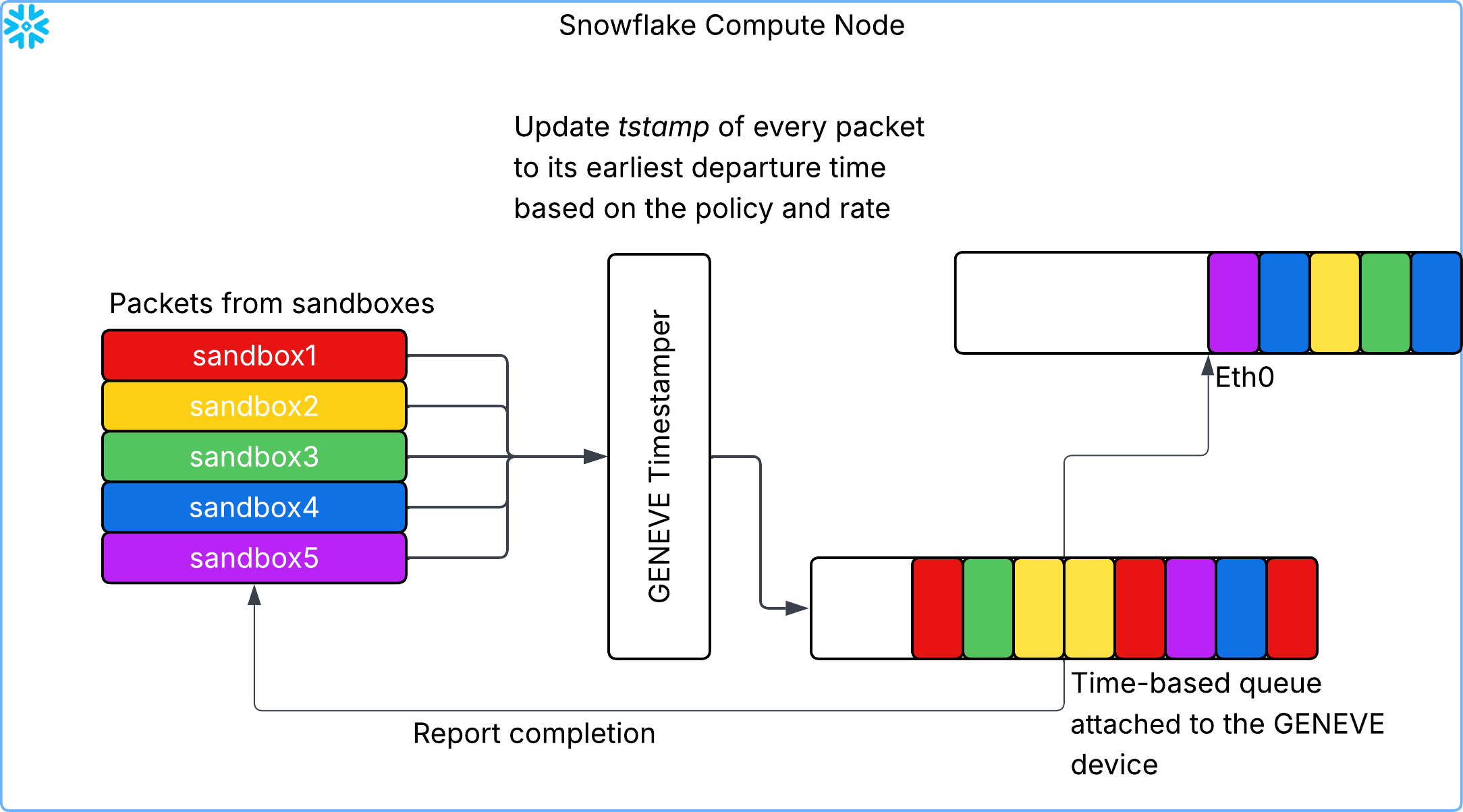}
      \captionof{figure}{EDT algorithm.}
      \label{fig:edt-alg}
      \vspace{0.2cm}
    \end{minipage}
    
    \item \textbf{Port Exhaustion Prevention.} Our overlay architecture inherently mitigates port exhaustion. Each sandbox operates in an isolated network namespace with independent and limited port space, and GENEVE encapsulation uses a single UDP port regardless of sandbox count. Additional safeguards include port range limitation at egress proxy, aggressive connection timeouts to free resources promptly, and per-sandbox connection limits to prevent runaway connection creation.
    
    \item \textbf{Connection Limiting for DDoS Prevention.} A hybrid sliding window algorithm limits requests at the sandbox level, combining fixed-window and sliding-window approaches. This prevents sustained abuse while accommodating legitimate burst patterns common in batch processing. Sandbox-level enforcement prevents malicious traffic from consuming compute nodes or proxy resources. The economic model makes attacks infeasible—launching attacks by spawning multiple sandboxes incurs proportional Snowflake compute costs, making it economically unattractive compared to traditional botnets.
\end{enumerate}

\subsection{Architecture Advantages}
The architecture achieves our design objectives through careful engineering choices. 

\begin{enumerate}
    \item \textbf{Security} is enforced through defense-in-depth with validation at compute nodes and proxy tiers. Cryptographically signed policies prevent tampering, and the zero-trust model assumes potential compute node compromise. 
    \item \textbf{Performance} is maintained through sub-millisecond eBPF packet processing. The kernel-space implementation eliminates context switching overhead, and EDT bandwidth limiting proves superior to TC and SQM alternatives. 
    \item \textbf{Scalability} comes from stateless GENEVE routing, distributed egress proxy architecture, and horizontal scaling across deployments. 
    \item \textbf{Transparency} is achieved as standard libraries work unmodified, user code remains portable, and no Snowflake-specific APIs are required.
\end{enumerate}

%% file: chapters/challenges.tex
\section{Challenges and Solutions}
\label{sec:challenges}

\subsection{eBPF Verifier Constraints}

\begin{figure}[ht!]
  \centering
  \includegraphics[width=0.9\linewidth]{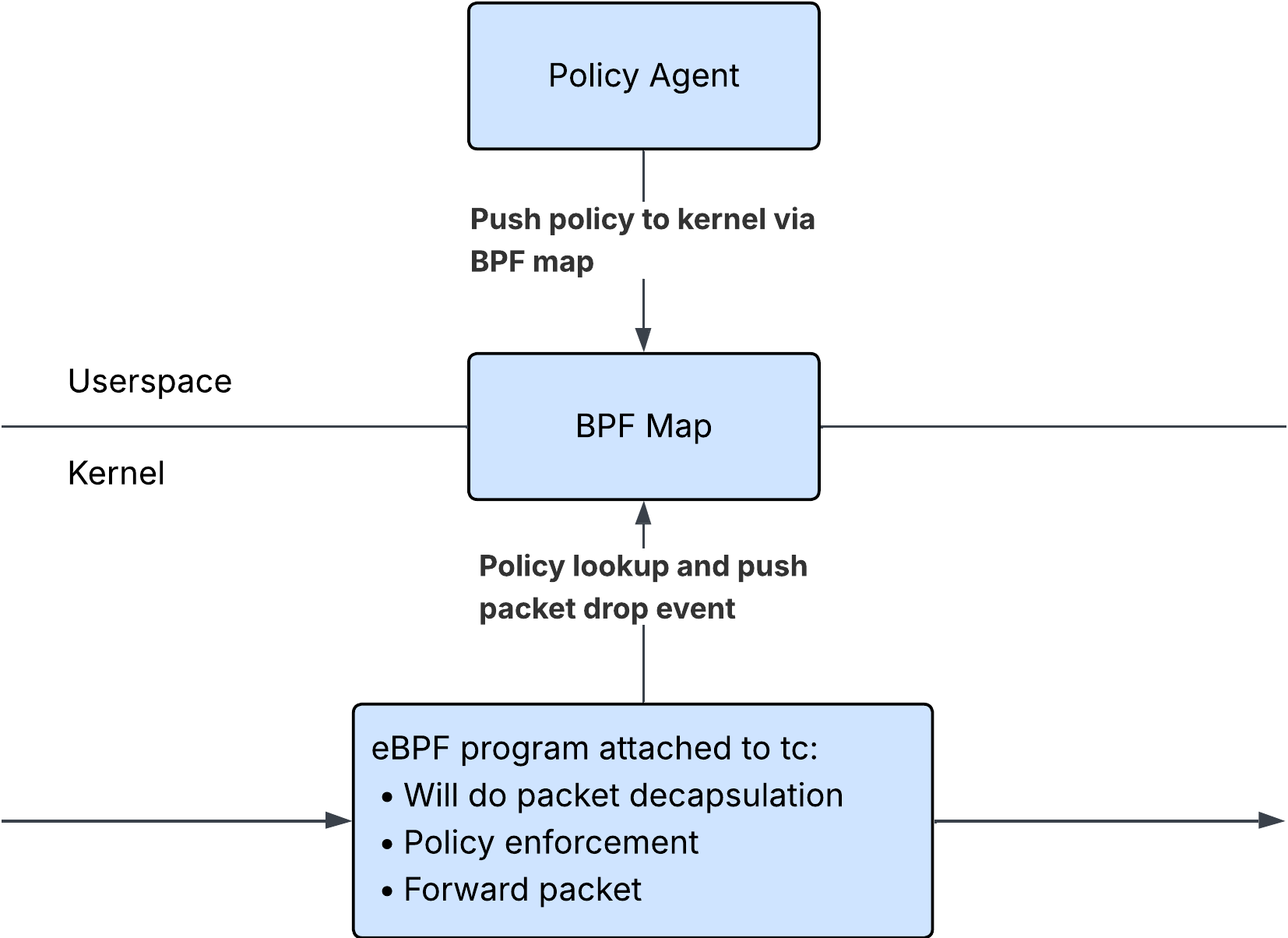}
  \caption{SNAS Policy Agent Logical Architecture.}
  \label{fig:policy-agent}
  \vspace{-0.3cm}
\end{figure}

The eBPF verifier enforces strict safety guarantees creating implementation challenges. Bounded loops require proving termination with explicit bounds checking. Stack size limits necessitated using eBPF maps for temporary storage. Complexity limits required decomposing functionality across multiple programs chained via tail calls~\cite{cite25}. We addressed these through modular eBPF design separating packet parsing, policy lookup, action enforcement, and logging into specialized programs, as outlined in Fig.~\ref{fig:policy-agent}.

\subsection{GENEVE Performance Optimization}

We successfully optimized GENEVE performance by enabling NIC hardware offload for checksum computation and segmentation (TCP Segmentation Offload / Generic Segmentation Offload (TSO/GSO)), which resulted in a significant reduction in CPU usage. For decapsulation, we implemented eXpress Data Path (XDP)-based eBPF processing pre-network-stack~\cite{cite23}, achieving additional overhead reduction through early packet processing before the packet reaches the full network stack.

\subsection{Egress Proxy Scalability}

High-cardinality workloads stressed connection tracking tables. We implemented distributed conntrack using consistent hashing—GENEVE packets route to proxy nodes based on hash of source IP, destination IP, and destination port, ensuring connection affinity~\cite{cite24}. Aggressive timeout policies prevent accumulation, with connections evicted after a brief, configurable idle period. Auto-scaling based on connection count and throughput metrics enables elastic capacity, with integration into Snowflake's resource orchestration enabling scaling within minutes.

\subsection{Resource Limitation Tuning}

The fine-grained control over resource management was achieved through a data-driven tuning process, leveraging extensive empirical analysis of production telemetry to deliver optimal and fair performance across diverse customer workload patterns. We analyzed bandwidth distribution, request rate distribution, and concurrent connection patterns. Limits were set at multiples of high percentiles to provide headroom while protecting infrastructure. Warehouse-specific tuning adjusts limits based on underlying infrastructure capacity. Customer feedback during preview validated the calibration approach, with minimal jobs hitting limits under normal operation.

\subsection{Security - Policy Integrity}

Policies must be tamper-proof to prevent compromised compute nodes from modifying them to allow unauthorized destinations. Policies are signed using Ed25519~\cite{cite10} with keys held exclusively by Cloud Services. Egress proxies validate signatures before loading into eBPF maps. Invalid signatures trigger immediate rejection and security alerts. Compute nodes include attestation metadata verified by proxies to ensure authorization for executing jobs for the associated account.

%% file: chapters/evaluation.tex
\section{Measured Operational Evaluation}
\label{sec:evaluation}

The system was evaluated using representative, large-scale production workloads, encompassing major data warehouse operations, transfer of extremely large datasets, processing of vast record counts, and high-volume external access patterns. Testing was performed on cloud infrastructure representative of the final deployment environments. 
Unless otherwise noted, all reported metrics are derived from production telemetry collected from live workloads running across multiple Snowflake cloud regions over two years of deployment. These measurements reflect field behavior of the deployed system.

\begin{figure}[ht!]
  \centering
  \vspace{-0.1cm} 
  \includegraphics[width=0.9\linewidth]{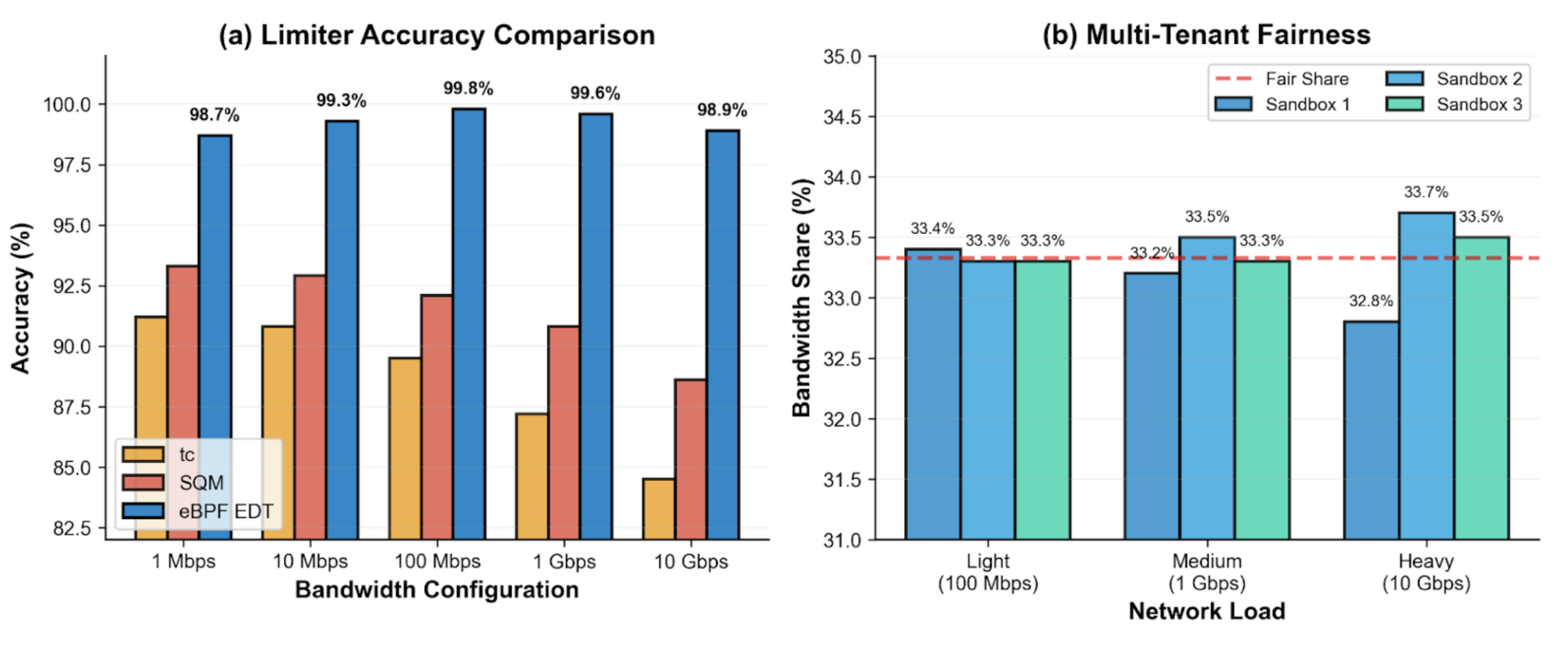}
  \caption{EDT bandwidth limiter accuracy and multi-tenant fairness. Left: achieved bandwidth vs. configured target across three rate settings. Right: per-tenant bandwidth distribution with concurrent sandboxes, demonstrating proportional fairness.}
  \label{fig:edt-accuracy}
\end{figure}

\begin{enumerate}
    \item \textbf{Latency Analysis:} measured via detailed kernel instrumentation~\cite{cite11}, we demonstrate that the packet processing overhead was confined to the sub-millisecond scale. This minimal processing overhead is negligible when compared against typical network round-trip times (RTTs), validating the efficiency of the kernel-space implementation. End-to-end request latency for same-region traffic showed the system's overhead is dominated by the physical network hop, with the eBPF datapath adding negligible latency ($<50\mu s$) to the total request time. Specifically, at p90 the eBPF enforcement path completed in single-digit microseconds; at p99, latency remained under $50\mu s$, with GENEVE encapsulation adding only a few additional microseconds. As a baseline comparison, a userspace proxy approach exhibited an order-of-magnitude higher per-request overhead due to context switching.

    \item \textbf{Throughput Analysis:} Single-connection throughput validated that the added encapsulation headers introduced negligible overhead, achieving over 90\% of NIC line rate through the full SNAS overlay. Critically, aggregate throughput scaled linearly as concurrent connections increased, demonstrating high stability and linear scalability under increasing load. Each egress proxy instance sustained high data transfer rates and concurrent session counts, confirming the system's robustness for demanding production environments.
    \item \textbf{Bandwidth Limiter Effectiveness:} The Earliest Departure Time (EDT)-based bandwidth limiter exhibited high accuracy across all tested operational thresholds, achieving ±2\% accuracy across the tested rates shown in Fig.~\ref{fig:edt-accuracy} with a coefficient of variation <3\% across concurrent tenants. Furthermore, the limiter achieved consistent fairness in resource distribution among multiple concurrent workloads. Comparative analysis against traditional queuing and shaping methods—including Linux TC token bucket filter (±8\% accuracy, CV >15\%) and userspace shaping (±12\% accuracy)—confirmed that the eBPF EDT approach achieved superior and more precise packet scheduling capabilities. This bandwidth limitation capability significantly reduced network-related resource conflicts after deployment, effectively protecting shared infrastructure.
\end{enumerate}

%% file: chapters/deployment.tex
\section{Deployment and Real-World Applications}
\label{sec:deployment}

\subsection{Global Production Scale}

SNAS is deployed across all global Snowflake regions as the egress mechanism for the platform's serverless compute layer. Over two years of production operation, it has served millions of daily jobs, including petabyte-scale transfers, while managing petabytes of aggregate egress traffic. The distributed proxy tier has proven resilient to individual proxy and zonal failures: each job uses a pool of proxy instances across availability zones, refreshed periodically by the control plane. If a proxy fails mid-connection, connections terminated on that proxy reset; clients reconnect and are routed to a healthy proxy via consistent-hash conntrack (Section~\ref{sec:challenges}). This fail-closed behavior makes proxy failures visible as ordinary connection retries rather than silent policy bypasses. Regarding security outcomes, Tier 1 has blocked substantial policy violation attempts, with Tier 2 serving as an independent backstop. To date, no policy violation has been observed to bypass both tiers simultaneously. This operational record is not a formal guarantee, but provides practical confidence in the defense-in-depth design.

\subsection{Workload Characterization}

SNAS deployment has revealed four workload patterns that validate our design:

\begin{enumerate}
    \item \textbf{Transparency for Connector Ecosystems:} A major barrier to adopting serverless ingestion is the need to adapt code to platform-specific service interfaces~\cite{cite17}. The transparency of our overlay network allows developers to run existing Python-based connectors (e.g., for CRM or financial data) directly within the sandbox. Because SNAS supports standard socket operations, these connectors function without modification, reducing the friction of ingesting data from third-party SaaS platforms.
    \item \textbf{Low-Latency Interactivity for AI Agents:} RAG workloads~\cite{cite18} require real-time access to external LLMs, vector stores, and web/API context sources; SNAS supports these interactions within interactive latency budgets.
    \item \textbf{Isolation for Regulated Logistics and IoT:} In shipping and pharmaceuticals, data often resides in regulated external environments. Per-sandbox namespaces and GENEVE encapsulation allow secure access to geospatial or sensitive vault APIs while cryptographically restricting connectivity to authorized endpoints.
    \item \textbf{Control Plane Orchestration:} Beyond data transfer, the platform is increasingly used as a control plane. Workloads utilizing the multi-protocol support (e.g., HTTP, gRPC) of SNAS can trigger external CI/CD pipelines or orchestration DAGs~\cite{cite19} upon data arrival. This effectively turns the data warehouse into an active participant in the broader enterprise infrastructure, capable of driving downstream automation.
\end{enumerate}

\subsection{Operational Lessons Learned}
SNAS at a global scale provided several key insights: 
\begin{enumerate}
    \item \textbf{Granularity Matters:} We found that FQDN-based allow-listing was significantly more popular than IP-based rules, validating the decision to handle secure DNS resolution within the control plane.
    \item \textbf{Resource Fairness:} The EDT-based bandwidth limiter proved essential not just for limiting abuse, but for preventing "noisy neighbor" contention during massive parallel data ingestion jobs.
    \item \textbf{Operational Simplicity Matters:} Keeping the datapath transparent to standard libraries and protocols reduced customer migration friction and lowered the need for workload-specific networking workarounds, which was important for production adoption at scale.
\end{enumerate}

%% file: chapters/conclusion.tex
\section{Conclusion}
\label{sec:conclusion}

SNAS is Snowflake's production approach to secure external connectivity for serverless data platforms. By combining eBPF packet filtering, GENEVE overlays, distributed egress proxies, and resource management, SNAS provides isolation, fairness, and developer transparency in a large multi-tenant deployment. Dual-tier policy enforcement, signed policies, and stateless GENEVE metadata provide operational controls with low datapath overhead. EDT bandwidth limiting and hybrid sliding-window rate limiting reduce noisy-neighbor and abusive traffic patterns. Because standard networking libraries work without Snowflake-specific APIs, workloads can adopt SNAS with low migration friction. Across all Snowflake regions, SNAS now supports millions of daily jobs and large-scale transfers, providing a practical template for secure external connectivity in multi-tenant cloud environments.

\textbf{Limitations and Future Work.} SNAS enforces IP:port policies and cannot inspect application-layer content such as HTTPS paths, headers, or payloads. This preserves end-to-end encryption but limits policy granularity to network-level controls. Policies are also job-scoped and static during execution. Future work includes Layer 7 filtering via eBPF socket-level hooks for finer-grained application-aware control.